\documentclass{article}
\usepackage{ijcai20}

\usepackage{times}
\usepackage{soul}
\usepackage{url}
\usepackage[hidelinks]{hyperref}
\usepackage[utf8]{inputenc}
\usepackage[small]{caption}
\usepackage{graphicx}
\usepackage{amsmath}
\usepackage{amsthm}
\usepackage{booktabs}
\usepackage{multirow, textcomp, xcolor}
\usepackage{algpseudocode}
\usepackage{subfigure}
\usepackage{pbox}

\usepackage{siunitx}

\graphicspath{{figures/}}
\newcommand\blfootnote[1]{%
  \begingroup
  \renewcommand\thefootnote{}\footnote{#1}%
  \addtocounter{footnote}{-1}%
  \endgroup
}

\title{Large Scale Audiovisual Learning of Sounds with Weakly Labeled Data}

\author{
    Haytham M.\ Fayek* and Anurag Kumar*\blfootnote{Equal contribution.}
    \affiliations
    Facebook Reality Labs, Redmond, WA, USA
    \emails
    \{haythamfayek, anuragkr\}@fb.com
}

\begin{document}

\maketitle

\begin{abstract}
	Recognizing sounds is a key aspect of computational audio scene analysis and machine perception.
	In this paper, we advocate that sound recognition is inherently a multi-modal audiovisual task in that it is easier to differentiate sounds using both the audio and visual modalities as opposed to one or the other.
	We present an audiovisual fusion model that learns to recognize sounds from weakly labeled video recordings.
	The proposed fusion model utilizes an attention mechanism to dynamically combine the outputs of the individual audio and visual models.
	Experiments on the large scale sound events dataset, AudioSet, demonstrate the efficacy of the proposed model, which outperforms the single-modal models, and state-of-the-art fusion and multi-modal models.
	We achieve a mean Average Precision (mAP) of 46.16 on Audioset, outperforming prior state of the art by approximately +4.35 mAP (relative: 10.4\%).
\end{abstract}

\section{Introduction}%
\label{sec:introduction}

Sound recognition is key for intelligent agents that perceive and interact with the world, similar to how the perception of different sounds is critical for humans to understand and interact with the environment.
Sounds are primarily produced by actions on or interactions between objects, and hence we often learn to immediately associate sounds with visual objects and entities.
This visual association is often necessary for recognizing and understanding the acoustic activities occurring around us.
For example, it is easier to distinguish between sounds produced by a \emph{Vacuum Cleaner} and sounds produced by a \emph{Hair Dryer} by listening to and seeing the appliances, as opposed to one or the other.

Sound recognition is inherently a multi-modal audiovisual task.
Thus, we ought to build machine learning models for sound recognition that are multi-modal, inspired by how humans perceive sound.
In addition to reducing uncertainties in certain situations as instantiated above, audiovisual learning of sounds can lead to a more holistic understanding of the environment.
For example, an alarm clock can be designed to produce \emph{Church Bell} sounds; while an audio only sound recognition model can perhaps label the sound as \emph{Church Bells}, which would not be incorrect, it nevertheless does not represent the actual event. Hence, systems designed to do audiovisual perception of sounds would lead to better and more complete understanding of these phenomena. 
However, most prior work on sound recognition focuses on learning to recognize sounds only from audio~\cite{virtanen2018computational,kumar2018acoustic}.

\emph{Multi-modal} learning approaches, which aim to train a single model by using the audio and visual modalities simultaneously are one class of well established audiovisual learning approaches. However, these methods present their own set of challenges. In these methods, it is often necessary to appropriately synchronize the audio and visual components and ensure temporal consistencies for a single model to consume both modalities.
Moreover, training a single multi-modal model which uses both modalities is a difficult feat due to different learning dynamics for the audio and visual modalities~\cite{Wang2019}.
Further, training and inference in multi-modal models is computationally cumbersome, particularly for large datasets~\cite{Gao2019}.

Another class of audiovisual learning methods can be described as \emph{Fusion} approaches. 
Fusion approaches aim to combine two independent single-modal models.
This offers several advantages.
We can build strong individual models for each modality separately, with the freedom to design modal-specific models without requiring to factor in the unnecessary complexities arising out of multi-modal models and training.
Furthermore, fusion approaches are often more interpretable than multi-modal models as it is easier to qualitatively analyze and understand the contributions of each modality through the individual models.
Finally, fusion approaches allow combining multiple models per modality, which can be advantageous in improving performance.

In this paper, we propose fusion based approaches for audiovisual learning of sounds. We propose methods that learn to combine individual audio and visual models that were respectively trained on each modality separately.
We first build state-of-the-art audio and visual sound recognition systems.
We then propose attention fusion models to dynamically combine these audio and visual models.
Specifically, our fusion models learn to pay attention to the appropriate modality in a \emph{sample-specific} \emph{class-specific} manner.
Our models are designed for weakly supervised learning, where we train models using weakly labeled data~\cite{Kumar2016}.
We analyze our proposed models on the largest dataset for sound events, Audioset~\cite{Gemmeke2017}, and show that the proposed models outperform state-of-the-art single-modal models, baseline fusion models, and multi-modal models.
The results and ensuing analysis attest to the importance of using both the audio and visual modalities for sound recognition and efficacy of our attention fusion models.

\section{Related Work}%
\label{sec:related}

\paragraph{Learning Sound Events.}
Numerous prior works have proposed a variety of supervised methods for detecting and classifying sound events~\cite{virtanen2018computational,kumar2018acoustic}.
Large-scale sound event detection has been possible primarily through weakly supervised learning~\cite{Kumar2016} and the release of large-scale weakly labeled sound events datasets, such as \emph{Audioset}~\cite{Gemmeke2017}.
Most of the recent methods rely on deep neural networks, in particular, deep Convolutional Neural Networks (ConvNets)~\cite{kumar2019secost,Adavanne2019,kong2019weakly,ford2019deep}.
While $mean$ and $max$ pooling have been shown to be effective in handling weakly labeled data~\cite{kumar2018acoustic}, more involved methods such as adaptive pooling~\cite{mcfee2018adaptive} or attention based pooling~\cite{wang2019comparison,kong2019weakly} have also been proposed.

\paragraph{Audiovisual Learning.}
There is a profusion of work on multi-modal audiovisual learning for video recognition and understanding in recent years~\cite{Baltruvsaitis2018}.
Progress in multi-modal learning was facilitated by the availability of large-scale video datasets~\cite{Kay2017,Gemmeke2017,Gu2018}, advances in self-supervised learning of multi-modal representations that exploit cross-modal information~\cite{Aytar2016,Arandjelovic2017,Owens2018,Korbar2018}, custom loss functions that synchronize learning multiple modalities~\cite{Wang2019}, and others.
Despite the recent progress, multi-modal audiovisual learning remains a challenging problem.
Primary reasons include difficulties in designing multi-modal neural network architectures that work well across various datasets and tasks, jointly learning multiple modalities, training and inference efficiency, etc.

\paragraph{Audiovisual Fusion.}
Late fusion approaches are a popular to combine audio and visual models~\cite{Ghanem2018}.
Unweighted late fusion such as averaging the outputs of the audio and visual models can demonstrate competitive performance in addition to not requiring any additional parameters and training beyond the audio and visual single-modal models~\cite{Lan2012}.
Nevertheless, weighted late or mid-level fusion has been extensively studied as it provides avenues to adaptively combine information between the different modalities~\cite{Lan2012,Wu2016,Kazakos2019}.
Further, attention mechanisms~\cite{Bahdanau2014} were used to dynamically combine the outputs of the single-modal models~\cite{Long2018,sterpu2018attention,hori2018multimodal,Zhou2019,Lin2019}.

\paragraph{Audiovisual Learning of Sounds}
Prior work on audiovisual learning of sounds, whether through single-model multi-modal learning or through fusion of individual models independently trained on the respective modalities, is scarce.
\cite{Wang2019} describes a multi-modal learning approach in which a single model is trained jointly for both the audio and visual modalities and applies this to sound recognition. 
Such methods are susceptible to several difficulties as outlined in Section~\ref{sec:introduction}.
We, on the other hand, propose fusion models for audiovisual learning of sounds and show in Section~\ref{tab:results} (Table~\ref{tab:art}) the effectiveness of our fusion models compared with the multi-modal models in~\cite{Wang2019}.
Another work related to ours is~\cite{parekh2019weakly}, which attempts to learn representations from weakly labeled audiovisual data and use these representations in different tasks including sound event recognition; therein, the audio and visual models are only combined through average fusion.
Perhaps worth mentioning are other works such as  as~\cite{Aytar2016,Arandjelovic2017,alwassel2019self}; they attempted to use the visual modality to learn representations for the audio modality that can be used for downstream sound event classification tasks.
However, in all prior work, the downstream sound classification tasks are done on small-scale datasets, which does not provide in-depth understanding of audiovisual learning for sounds at scale.

\section{Audiovisual Models for Sounds}%
\label{sec:audvis}

We learn sound events in the weakly supervised setting, where only recording-level labels are provided for each training recording, with no temporal information.
The fundamentals of weakly supervised learning of sound events are based on Multiple Instance Learning (MIL)~\cite{Kumar2016}.
In MIL, the learning problem is formulated in terms of \emph{bags} and respective labels, $(\mathcal{B}, Y)$;
each {\em bag} is a collection of instances~\cite{andrews2003support}.
A bag is labeled as positive for a class, $Y = 1$, if at least one instance within the bag is positive.
On the other hand, if all instances within the {\em bag} are negative for a given class, then the {\em label} is negative, $Y = 0$.
For weakly labeled sound recognition, recordings which are tagged with the presence of a sound class become a positive bag for that class and negative otherwise.
The audio and visual models in Sections~\ref{ssec:audio} and~\ref{ssec:visual} follow this formulation.

\subsection{Audio Model}%
\label{ssec:audio}
In the MIL setting for weakly labeled sound event recognition, each recording becomes a {\em bag} as required in the MIL framework.
The instances in the {\em bag } are segments of the recording.
Assuming that each instance is represented by a feature vector $\mathbf{x}$;
the training data takes the form of $(\mathcal{B}_i, Y_i),\:\:i=1$ to $n$, where the $i^{th}$ bag consists of $m_i$ instances $\mathcal{B}_i = \{\mathbf{x}_{i1} \cdots \mathbf{x}_{im_i}\}$.

In the audio model, the key idea is that the learner can be trained by first mapping the instance-level predictions to the corresponding bag-level predictions, and then these bag-level predictions can be used to compute a loss function to be minimized, akin to the classic MI-SVM approach for MIL~\cite{andrews2003support}.
Herein, the $max$ function was used to aggregate the instance-level predictions.

Let $f$ be the function we wish to learn;
then the training involves minimizing the following loss function.
\begin{equation}%
	\label{eq:trainloss}
	\mathcal{L}(\Theta, \Phi) = \sum_{i=1}^n l(g_{\Phi}(f_\Theta(\mathbf{x}_{i1}), \cdots, f_\Theta(\mathbf{x}_{im_i})), Y_i)
\end{equation}
where, $l$ is the loss function measuring divergence between predictions and targets, and $\Theta$ represents learnable parameters of $f$.
Note that the function $g$, used to aggregate the instance-level predictions from $f$ to the bag-level predictions, can itself contain learnable parameters $\Phi$.

The audio model is a ConvNet that maps Log-scaled Mel-filter-bank feature representations of the entire audio recording to (multiple) labels, trained by minimizing the loss function defined in Equation~\ref{eq:trainloss}.
The sampling rate of all audio recordings is \SI{16}{kHz}; $64$ Mel-filter-bank representations are obtained for \SI{16}{ms} windows in the audio recording, shifted by \SI{10}{ms}, which amounts to 100 frames per second of audio.

The overall architecture of the network is detailed in Table~\ref{tab:cnnarch}. The network produces {\em instance} and {\em bag} representations for any given input at Block B5. The bag consists of $2048$-dimensional instance representations. 
The network is designed for a receptive field of \texttildelow$1$ second ($96$ frames). Hence, it outputs $2048$-dimensional features for \texttildelow$1$ second of audio, every \texttildelow$0.33$ seconds ($32$ frames).
Three $1 \times 1$ convolutional layers (B6 to B8) are then used to predict outputs for each segment, which are then aggregated via the layer $G$ that uses global average pooling.
The number of segments depends on the size of the input.
As shown in Table~\ref{tab:cnnarch}, for an input with $1024$ frames, $30$ segments are produced.

The loss function for the $i^{th}$ training audio recording is defined as:
\begin{equation}%
	\label{eq:lossfn}
	l(\mathcal{B}_i,Y_i) = \frac{1}{C} \sum_{c=1}^C h(Y_c^i,p_c^i)
\end{equation}
where, $p_c^i$ is the output for the $c^{th}$ class and $h(Y_c^i,p_c^i) = -Y_c^i*\log(p_c^i) - (1-Y_c^i)*\log(1-p_c^i)$ is the binary cross entropy loss function.
Furthermore, our audio model training also incorporates the sequential co-supervision method described in~\cite{kumar2019secost}.

\subsection{Visual Model}%
\label{ssec:visual}

Similar to the audio model, the visual model for sounds is based on the MIL framework.
Herein, the entire video is the bag and instances of the bag are the frames of the video.
Specifically, we sample $64$ frames from the videos to form the bags.
Each frame (instance) is then represented by $2048$-feature representations obtained from a ResNet-152~\cite{He2016}, pre-trained on Imagenet~\cite{Deng2009}.
This yields the video bag representations for each recording.
We obtain a single $2048$-dimensional vector representation for each bag by averaging the feature representations for each instance.
This idea of mapping bags to a single vector representations was previously employed in~\cite{kumar2016weaklysupervised,wei2014scalable} to make MIL algorithms scalable.

We then train a $4$ hidden layer Deep Fully Connected Neural Network (DNN) to recognize sound events.
The number of units in each layer are $2048, 2048, 1024, 1024, C$ respectively, where $C$ is the number of classes.
The first and second hidden layers are followed by a dropout layer with a dropout rate of $0.3$.
Rectified Linear Units (ReLU) activation is used in all layers except for the last layer, where sigmoid activation is used. Similar to the audio model, the network is trained with the binary cross entropy loss function.

\begin{table}[t]
	\centering
	\resizebox{1.0\columnwidth}{!}{
		\begin{tabular}{c|c|c}
			\toprule
			\textbf{Stage}            & \textbf{Layers}                                 & \textbf{Output Size}       \\
			\midrule
			Input                     & Unless specified -- (S)tride = 1, (P)adding = 1 & $1 \times 1024 \times 64$  \\
			\midrule
			\multirow{3}{*}{Block B1} & Conv: 64, $3 \times 3$                          & $64 \times 1024 \times 64$ \\
			                          & Conv: 64, $3 \times 3$                          & $64 \times 1024 \times 64$ \\
			                          & Pool: $4 \times 4$ (S:4)                        & $64 \times 256 \times 16$  \\
			\midrule
			\multirow{3}{*}{Block B2} & Conv: 128, $3 \times 3$                         & $128 \times 256 \times 16$ \\
			                          & Conv: 128, $3 \times 3$                         & $128 \times 256 \times 16$ \\
			                          & Pool: $2 \times 2$ (S:2)                        & $128 \times 128 \times 8$  \\
			\midrule
			\multirow{3}{*}{Block B3} & Conv: 256, $3 \times 3$                         & $256 \times 128 \times 8$  \\
			                          & Conv: 256, $3 \times 3$                         & $256 \times 128 \times 8$  \\
			                          & Pool: $2 \times 2$ (S:2)                        & $256 \times 64 \times 4 $  \\
			\midrule
			\multirow{3}{*}{Block B4} & Conv: 512, $3 \times 3$                         & $512\times 64 \times 4 $   \\
			                          & Conv: 512, $3 \times 3$                         & $512 \times 64 \times 4 $  \\
			                          & Pool: $2 \times 2$ (S:2)                        & $512 \times 32 \times 2 $  \\
			\midrule
			Block B5                  & Conv: 2048, $3 \times 2$ (P:0)                  & $2048 \times 30 \times 1 $ \\
			\midrule
			Block B6                  & Conv: 1024, $1 \times 1$                        & $1024 \times 30 \times 1 $ \\
			\midrule
			Block B7                  & Conv: 1024, $1 \times 1$                        & $1024 \times 30 \times 1 $ \\
			\midrule
			Block B8                  & Conv: C, $1 \times 1$                           & $C \times 30 \times 1 $    \\
			\midrule
			G ($g()$)                 & Global average pooling                          & $C \times 1 $              \\
			\bottomrule
		\end{tabular}}
	\caption{Audio Convolutional Neural Network (ConvNet) architecture.
		The number of filters and size of each filter varies between convolutional layers; e.g., Conv: 64, $3 \times 3$ denotes a convolutional layer of 64 filters, each of size $3 \times 3$.
		$C$ denotes the number of classes.
		All convolutional layers except in B8 are followed by batch normalization and Rectified Linear Units (ReLU);
		the convolutional layer in B8 is followed by a sigmoid function.}%
	\label{tab:cnnarch}
\end{table}%

\begin{figure*}[t]
	\centering
	\includegraphics[width=0.9\linewidth]{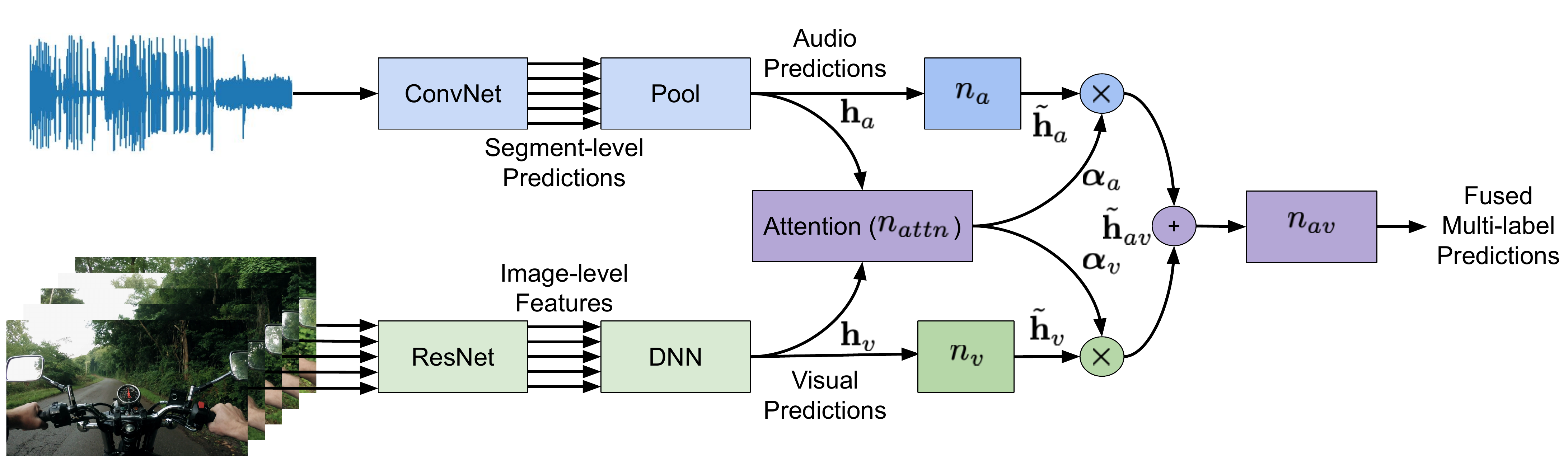}
	\caption{Audiovisual fusion for sound recognition.
		The audio and visual models map the respective input to segment-level representations which are  then used to obtain single-modal predictions, $\mathbf{h}_a$ and $\mathbf{h}_v$, respectively.
		The attention fusion function, $n_{attn}$, ingests the single-modal predictions, $\mathbf{h}_a$ and $\mathbf{h}_v$, to produce weights for each modality, $\boldsymbol{\alpha}_a$ and $\boldsymbol{\alpha}_v$.
		The single-modal audio and visual predictions, $\mathbf{h}_a$ and $\mathbf{h}_v$, are mapped to $\tilde{\mathbf{h}}_{a}$ and $\tilde{\mathbf{h}}_{v}$ via functions, $n_{a}$ and $n_{v}$, respectively, and fused using the attention weights, $\boldsymbol{\alpha}_a$ and $\boldsymbol{\alpha}_v$.
		The fused output, $\tilde{\mathbf{h}}_{av}$, is then mapped to multi-modal multi-label predictions via function $n_{av}$.}%
	\label{fig:fusion}
\end{figure*}

\subsection{Baseline Audiovisual Fusion}%
\label{ssec:sfusion}

Our audiovisual learning of sounds focuses on developing fusion models that combine outputs from models trained individually on the audio and visual modalities.
The following is a description of a number of baseline fusion methods.

\paragraph{Average Fusion.}
We simply compute the arithmetic mean of the outputs of the audio and visual models.
Despite its simplicity, average fusion serves as a strong baseline.
\cite{parekh2019weakly} used average fusion to combine audio and visual predictions.

\paragraph{Regression.}
We train a linear layer with a sigmoid activation function that ingests the (mean-standard-deviation) normalized outputs of the audio and visual models, and predicts the probability of the sound classes.
The model is $L2$ regularized with the regularization parameter set to $\lambda = 10^{-5}$.

\paragraph{Multi-Layer Perceptron (MLP).}
We train a fully connected neural network with a single hidden layer that ingests concatenated outputs of the audio and visual models, and predicts the probability of the sound classes.
The hidden layer comprises $512$ units with batch normalization and ReLU activation.
Furthermore, a dropout layer with a dropout rate of $0.5$ is used.
The output layer is a fully connected layer with $C$ units and a sigmoid activation function.

\subsection{Attention Audiovisual Fusion}%
\label{sec:afusion}

The audio and visual models have been separately trained to map the audio and visual modalities, respectively, to the various sound classes.
The performance of each class varies between modalities.
%
This variation can be sample-specific as well.
For a given sound class, in some videos, the audio modality might be sufficient to correctly recognize the sound, whereas in other videos, the visual modality might be more suited.
Thus, we ought to fuse the outputs from the audio and visual models in a sample-specific and class-specific manner.
To this end, we design an attention mechanism to combine the audio and visual models.

Figure~\ref{fig:fusion} shows the general schema of the attention fusion mechanism.
Let $\mathbf{h}_a$ and $\mathbf{h}_v$ represent the outputs of the audio and visual models for a given video recording respectively.
We learn an attention function, $a(\cdot)$, that generates weights $\boldsymbol{\alpha}_a$ and $\boldsymbol{\alpha}_v$ for the audio and visual modalities respectively.
We constrain the attention weights to sum to $1$ for each class, i.e., $\boldsymbol{\alpha}_a \,+\, \boldsymbol{\alpha}_v \,=\, \mathbf{1}$, where $\mathbf{1}$ is a vector of ones.

The attention function is implemented as a neural network, $n_{attn}$:
\begin{equation}%
	\label{eq:attnwt}
	\boldsymbol{\alpha}_a = n_{attn}([\mathbf{h}_a, \mathbf{h}_v], W_{attn})
\end{equation}
where, $W_{attn}$ denotes the learnable parameters of network $n_{attn}$, and
$[\mathbf{h}_a, \mathbf{h}_v]$ denotes the concatenation of $\mathbf{h}_a$ and $\mathbf{h}_v$.
The output layer in $n_{attn}$ has a sigmoid activation function.

The final output that combines the audio and visual models is obtained as follows:
\begin{gather}
	\tilde{\mathbf{h}}_a =  n_a(\mathbf{h}_a, W^n_a)\:\: \text{,} \:\: \tilde{\mathbf{h}}_v =  n_v(\mathbf{h}_v, W^n_v) \label{eq:finalo1} \\
	\tilde{\mathbf{h}}_{av} =  \boldsymbol{\alpha}_a \odot \tilde{\mathbf{h}}_a + (1 -  \boldsymbol{\alpha}_a) \odot \tilde{\mathbf{h}}_v \label{eq:finalo2} \\
	\mathbf{o}_{av} = n_{av}(\tilde{\mathbf{h}}_{av}, W^n_{av}) \label{eq:finalo3}
\end{gather}
where, $n_a$ and $n_v$ are neural networks that process inputs $\mathbf{h}_a$ and $\mathbf{h}_v$ into $\tilde{\mathbf{h}}_a$ and $\tilde{\mathbf{h}}_v$, respectively, which are then combined through attention weights ($\boldsymbol{\alpha}_a$ and $\boldsymbol{\alpha}_v = 1 - \boldsymbol{\alpha}_a$ ), as in Equation~\ref{eq:finalo2}.
$W^n_a$ and $W^n_v$ are the learnable parameters of $n_a$ and $n_v$ respectively.
The combined outputs, $\tilde{\mathbf{h}}_{av}$, are passed through another neural network $n_{av}$ with a sigmoid activation function to produce the final outputs $\mathbf{o}_{av}$.

Note that $n_a$ and $n_v$ are generic functions representing all possible transforms: linear, non-linear, or even the identity of $\mathbf{h}_a$ and $\mathbf{h}_v$ respectively.
For the identity case, $n(\mathbf{h}) = \mathbf{h}$, and $n$ does not contain any learnable parameters.
For the linear transform case, $n$ does not contain any non-linear activation functions.
Similarly, $n_{av}$ might just be a sigmoid function, i.e., $n_{av}(\tilde{\mathbf{h}}_{av}) = 1/(1 + \exp^{-\tilde{\mathbf{h}}_{av}})$, and hence, containing no learnable parameters.
Otherwise, it can be a neural network with one or more layers with a sigmoid activation at the output layer.
We also attempted to learn the attention weights with a single modality, either $\mathbf{h}_a$ or $\mathbf{h}_v$ instead of the combined $[\mathbf{h}_a, \mathbf{h}_v]$.
In this case, $\boldsymbol{\alpha}_a$ is either obtained through self-attention $\boldsymbol{\alpha}_a = n_{attn}(\mathbf{h}_a, W_{attn})$ or through cross-modal attention $\boldsymbol{\alpha}_a = n_{attn}(\mathbf{h}_v, W_{attn})$.
Similarly, $\boldsymbol{\alpha}_v$ is obtained, and both $\boldsymbol{\alpha}_a$ and $\boldsymbol{\alpha}_v$ are normalized to sum to $1$ for each class.
However, attention weights learned from single modality was found to be inferior to learning them through both modalities as in Equation~\ref{eq:attnwt}.

\section{Experiments}%
\label{sec:experiments}

Experiments are conducted on AudioSet which is described in Section~\ref{ssec:dataset}.
The experiments and results are presented in Section~\ref{ssec:baseline}.

\subsection{Audioset Dataset}%
\label{ssec:dataset}

Audioset~\cite{Gemmeke2017} is the largest dataset for sound events.
The dataset provides YouTube videos for $527$ sound events.
Each video clip is approximately \SI{10}{s} in length annotated by a human with multiple labels denoting the sound events present in the video clip.
The average number of labels per video clip is 2.7.
The dataset is weakly labeled as the labels of each video clip denote the absence or presence of sound events but do not contain any temporal information.
The distribution of sound event classes in the training set is severely unbalanced, ranging from around $1$ million videos for the most represented class, \emph{Music}, to approximately $120$ videos for the least represented class, \emph{Screech}.

We use the predefined {\em Unbalanced} set for training and {\em Eval} set for performance evaluation.
The training set comprises approximately $2$ million videos, whereas the evaluation set comprises approximately $20,000$ videos.
The evaluation set contains at least $59$ videos for each class.
In practice, the total video count is slightly less than the numbers listed above due to the unavailability of links when this work was carried out.
We sample approximately $25,000$ videos from the training set to use as the validation set.

\subsection{Performance Evaluation of Fusion Methods}%
\label{ssec:baseline}

\paragraph{Experimental Setup.}
The outputs (predictions) of the audio and visual models are used as input to all fusion models.
These outputs are normalized to zero mean and unit standard deviation prior to feeding them into the fusion models (except for average fusion).
The mean and standard deviation are computed using the training set only.

The details of the audio model training can be found in~\cite{kumar2019secost}. 
The network in the video model is trained for 20 epochs using the Adam optimizer~\cite{Kingma2014} with a mini-batch size of 144. Hyperparameters such as the learning rate are selected using the validation set, which is also used to select the best model during training. 

In the fusion experiments, all neural networks are trained using Adam for 100 epochs.
The mini-batch size is set to $256$.
The validation set is used to select the best hyperparameters, such as the learning rate for each model, and the best model during training.
Similar to prior work on Audioset, we use Average Precision (AP) and Area Under Receiver Operating Characteristic (ROC) Curve (AUC) to measure the performance per class.
Mean Average Precision (mAP) and mean Area Under ROC Curve (mAUC) computed over all $527$ classes are used for comparison.

\begin{table}[t]
	\centering
	\begin{tabular}{l|c|c}
		\toprule
		Model                     & mAP            & mAUC           \\
		\midrule
		Audio                     & 38.35          & 97.12          \\
		Visual                    & 25.73          & 91.30          \\
		\midrule
		Average Fusion            & 42.84          & \textbf{97.63} \\
		Regression Fusion         & 43.10          & 95.35          \\
		MLP Fusion                & 45.60          & 96.83          \\
		\midrule
		\textbf{Attention Fusion} & \textbf{46.16} & 97.51          \\
		\bottomrule
	\end{tabular}
	\caption{Comparison of mAP and mAUC for different fusion methods for combining audio and visual models.}%
	\label{tab:results}
\end{table}

\begin{table}[t]
	\centering
	\resizebox{1.0\columnwidth}{!}{
		\begin{tabular}{l|c|c}
			\toprule
			Model                                  & mAP            & mAUC           \\
			\midrule
			Audio (ResNet-50) ~\cite{ford2019deep} & 38.0 & 97.10          \\
			Audio (Ours)                           & \textbf{38.35}          & \textbf{97.12} \\
			\midrule
			Visual (R(2+1)D-101) ~\cite{Wang2019}  & 18.80          & \textbf{91.80} \\
			Visual (Ours)                          & \textbf{25.73} & 91.30          \\
			\midrule
			AudioVisual (G-Blend) ~\cite{Wang2019} & 41.80          & 97.50          \\
			\textbf{AudioVisual (Ours)}            & \textbf{46.16} & \textbf{97.51} \\
			\bottomrule
		\end{tabular}
	}
	\caption{mAP and mAUC for state-of-the-art audio, visual, and audiovisual sound recognition models on AudioSet.}%
	\label{tab:art}
\end{table}

\paragraph{Comparison of Fusion Methods.}
Table~\ref{tab:results} summarizes the results.
The audio model achieves 38.35 mAP and 97.12 mAUC, whereas the visual model achieves 25.73 mAP and 91.30 mAUC.
The superiority of the audio model over the visual model is expected due to the nature of the task.

Average fusion of the audio and visual outputs achieves 42.84 mAP, an absolute improvement of 4.49 mAP (relative: 11.7\%) over the audio model, and an absolute improvement of 17.11 mAP (relative: 66.5\%) over the visual model.
The regression fusion model leads to a minor improvement over average fusion: +0.26 mAP.
The MLP fusion model leads to considerable improvement over average fusion: +2.76 mAP.

Our attention fusion model achieves 46.16 mAP, which is an absolute improvement of 7.81 mAP (relative: 20.4\%) over the audio model.
It also outperforms all baseline fusion methods: an absolute improvement of 3.32 mAP (relative: 7.7\%) over average fusion.
In this case, $n_{attn}$ takes in the concatenated normalized outputs from the audio ($\mathbf{h}_a$) and visual ($\mathbf{h}_v$) models.
$n_{attn}$ is a neural network with a single hidden layer of 512 units;
batch normalization and ReLU activation is applied on this layer which is followed by a dropout layer with a dropout rate of 0.5.
The output layer has 527 units and a sigmoid activation function. 
$n_a$, $n_v$, and $n_{av}$ are all single layer networks with 512 units and sigmoid activations.

We also study different variations of the proposed fusion model.
The \emph{unimodal} approaches uses only a single modality in the attention network, either $\mathbf{h}_a$ or $\mathbf{h}_v$ as opposed to using $\mathbf{h}_a$ and $\mathbf{h}_v$ combined through concatenation.
When $\boldsymbol{\alpha}_a$ is obtained by using $\mathbf{h}_a$ and $\boldsymbol{\alpha}_v$ using $\mathbf{h}_v$, we denote it as \emph{self unimodal} attention.
However, if $\boldsymbol{\alpha}_a$ uses $\mathbf{h}_v$ and vice versa, we call it \emph{cross unimodal} attention.
In these cases, there are two attention networks, $n_{attn}^a$ and $n_{attn}^v$ but other aspects of the architecture remains identical to the one described above.

We also consider another variation where $n_a$ and $n_v$ are simply identity functions that do not contain learnable parameters.
Similarly, in another variation, we make $n_{av}$ an identity mapping.
The performance of these fusion models are listed in Table~\ref{tab:ablation}.
The multi-modal attention model is slightly better than the unimodal attention models.
However, removing neural networks $n_a$, $n_v$, and $n_{av}$ leads to significant deterioration in performance.

\begin{table}[t]
	\centering
	\begin{tabular}{l|c|c}
		\toprule
		Model                                       & mAP            & mAUC           \\
		\midrule
		Attention (self unimodal)                   & 46.04          & 97.48          \\
		Attention (cross unimodal)                  & 46.06          & 97.47          \\
		Attention (multi-modal w/o $n_a$ and $n_v$) & 38.84          & 95.34          \\
		Attention (multi-modal w/o $n_{av}$)        & 40.84          & 95.15          \\
		\textbf{Attention (multi-modal)}            & \textbf{46.16} & \textbf{97.51} \\
		\bottomrule
	\end{tabular}
	\caption{Ablation of attention fusion models.}%
	\label{tab:ablation}
\end{table}

\paragraph{Comparison with State-of-the-Art.}
Table~\ref{tab:art} shows comparison with state-of-the-art models on Audioset.
Our audio model is slightly better than the state-of-the-art performance on Audioset. Note that \cite{ford2019deep} also shows a performance of 39.2 mAP. However, this is obtained by a ensembeling outputs from multiple models and the best single model performance (which is fairer to compare with) is 38.0. To the best of our knowledge,~\cite{Wang2019} is the only prior work that reported visual and audiovisual models for sound events on Audioset.
Our visual model is +6.93 mAP (relative: 36.8\%) better than~\cite{Wang2019}.
More importantly, our audiovisual model is \textbf{+4.35 mAP (relative: 10.4\%)} compared with~\cite{Wang2019} and sets a new state-of-the-art on Audioset.

\section{Discussions}%
\label{sec:discussion}

\begin{figure}[t]
	\centering
	\includegraphics[width=\linewidth]{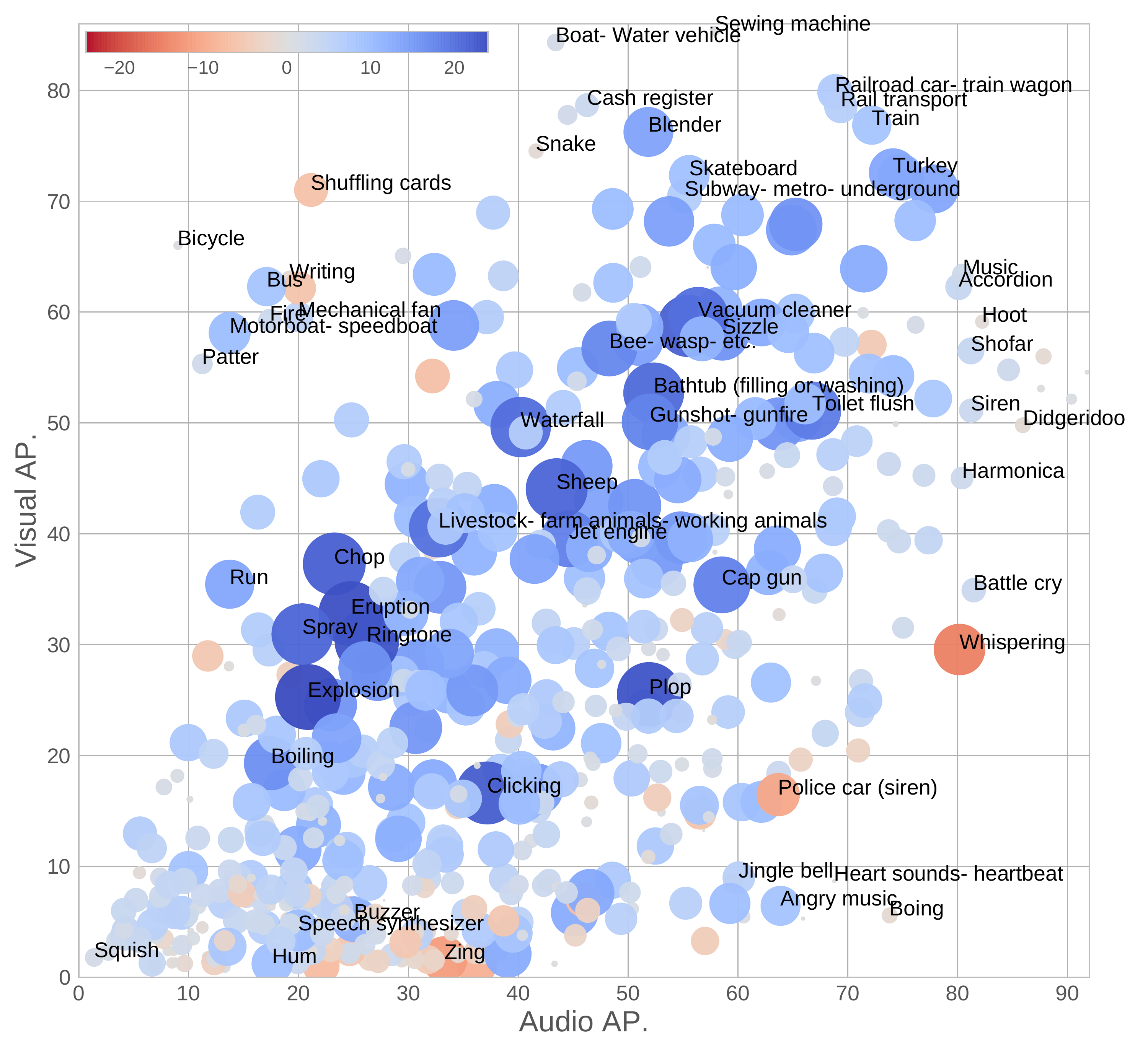}
	\caption{Audio, visual, and audiovisual models Average Precision (AP). The horizontal and vertical axes denote the audio and visual models APs respectively for all classes. The diameter of each bubble denotes the absolute change in AP over the single-modal models (audio or visual, whichever is better). Blue denotes improvement over the single-modal model, whereas red indicates deterioration over the single-modal model, as depicted in the color bar. Several sound classes of interest are annotated accordingly.}%
	\label{fig:classmap}
\end{figure}

\subsection{Class Specific Analysis}
Figure~\ref{fig:classmap} provides a per class depiction of the audio, visual, and audiovisual models.
As expected, the audio model outperforms the video model for the majority of sound events.
For sounds such as \emph{Heartbeat}, \emph{Jingle Bell}, and \emph{Whispering}, audio recognition of sounds is considerably better than visual recognition of sounds.
However, for classes such as \emph{Patter}, \emph{Shuffling Cards}, and \emph{Bus}, the visual model outperforms the audio model by a large margin.
Overall, the visual model outperforms the audio model for 117 sound classes by a small or large margin.
For some classes, both audio and visual models perform well (e.g., \emph{Turkey}, \emph{Train}, and \emph{Music Accordion}), or both fail to achieve good performance (e.g., \emph{Squish} and \emph{Hum}).

\begin{figure}[t]
	\centering
	\includegraphics[width=0.8\linewidth]{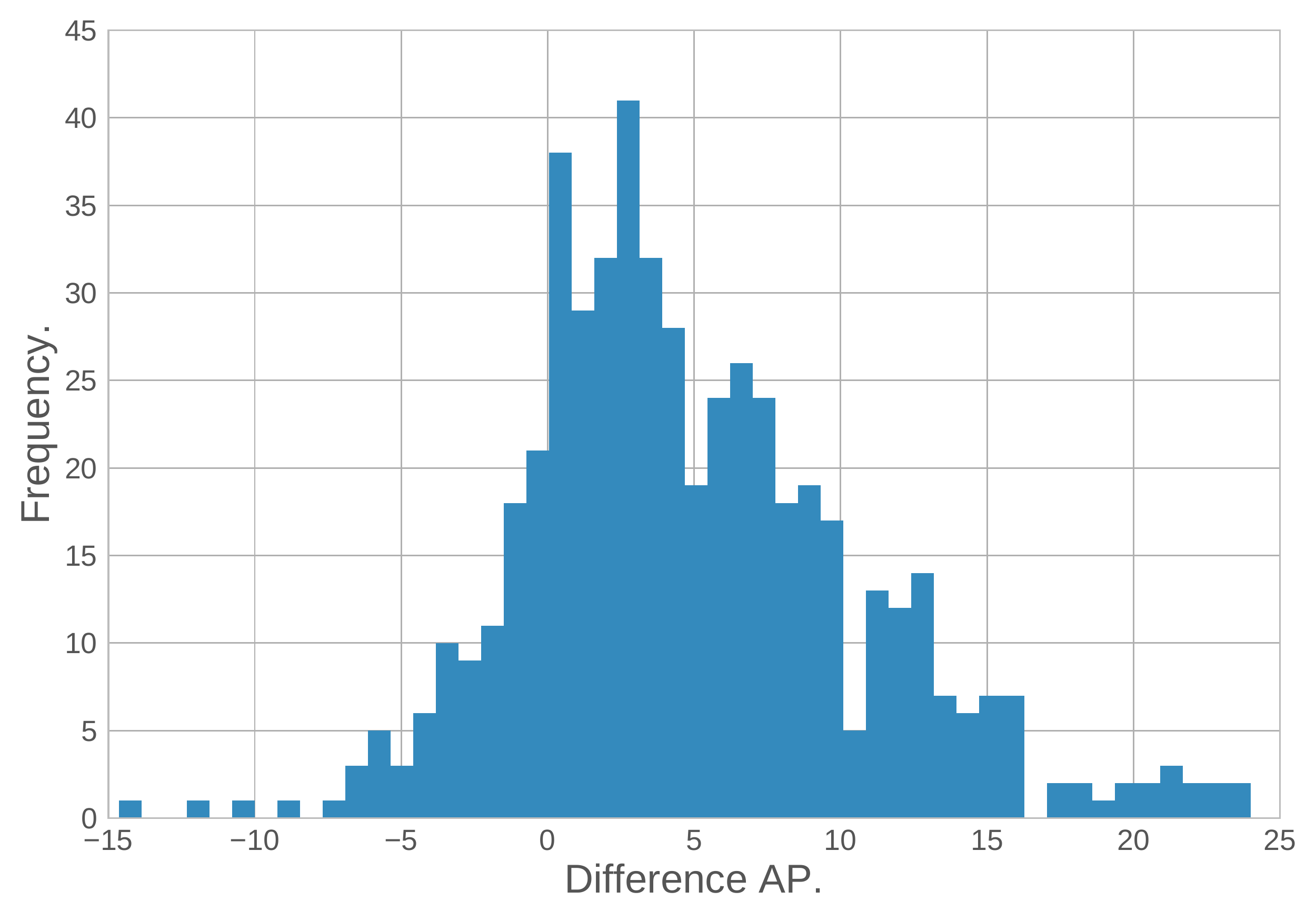}
	\caption{Histogram of change in AP for our audiovisual attention fusion model with respect to single modality models (audio or visual whichever is better).}%
	\label{fig:improvement}
\end{figure}

For the audiovisual fusion model, we see that for most classes, audiovisual learning leads to an improvement over the better of the audio or visual models, as observable by mostly blue bubbles in Figure~\ref{fig:classmap}.
Figure~\ref{fig:improvement} shows a histogram of the absolute deterioration and improvement in performance through audiovisual learning compared to the better of the audio or visual models.
For over 83\% of classes (439 of 527), we see an improvement, whereas for the remaining classes, audiovisual learning leads to deterioration in performance, compared with the better of the audio or visual models.
Furthermore, for over 17\% of classes (91 out of 527), audiovisual learning achieves an absolute improvement of more than +10 mAP.
On the other hand, for only 15 classes audiovisual learning leads to an absolute deterioration of more than 5 mAP, with only 3 classes with more than 10 deterioration in mAP.
For most cases, the deterioration is minor.

\begin{figure}[t]
	\centering
	\includegraphics[width=0.8\linewidth]{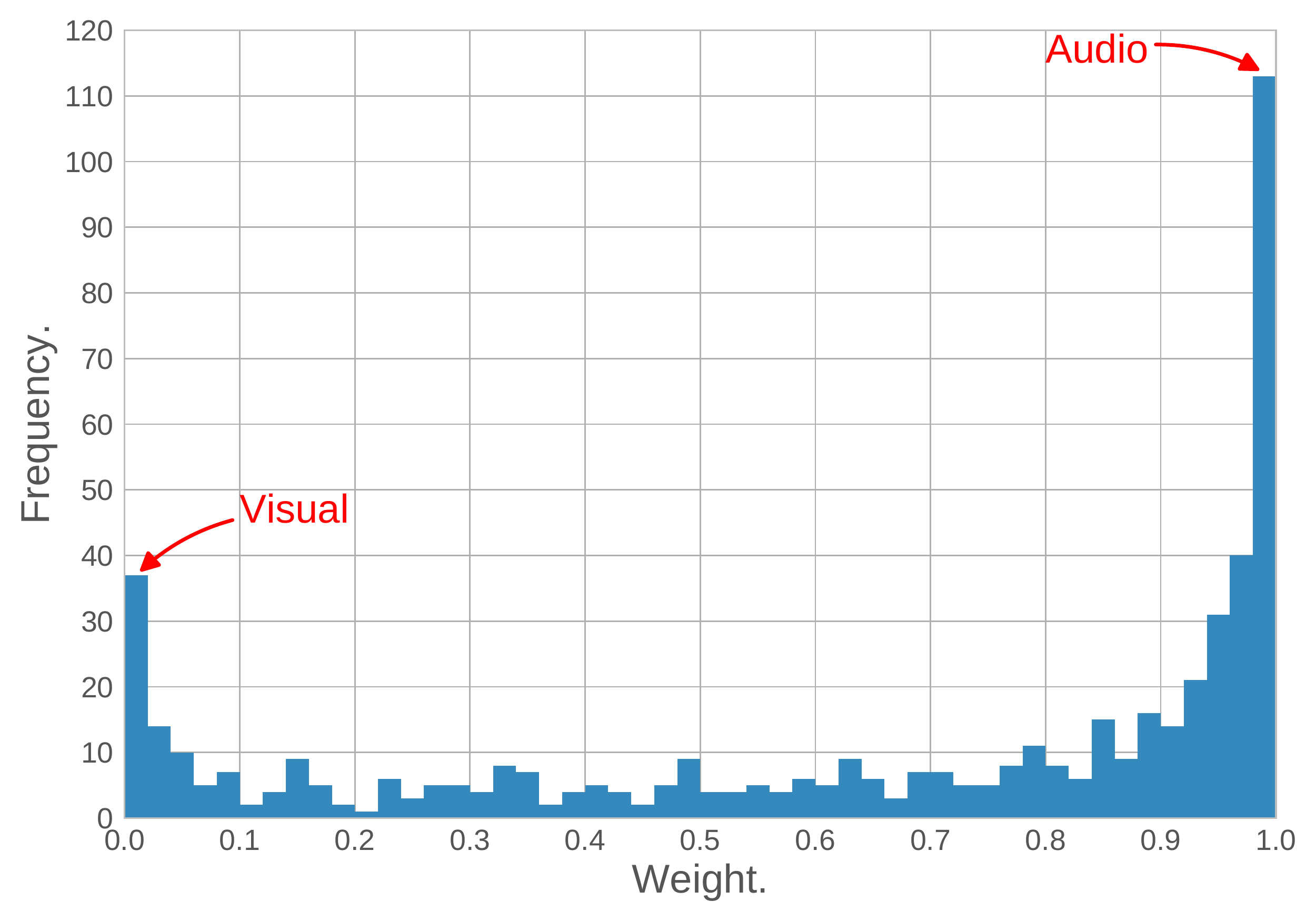}
	\caption{Mean of attention weights for all classes in AudioSet.}%
	\label{fig:weights}
\end{figure}

From Figure~\ref{fig:classmap}, we can see some examples of classes with large improvements in the audiovisual model, e.g., \emph{Explosion}, \emph{Ringtone Clicking}, \emph{Vacuum Cleaner}, and \emph{Bathtub}.
Several of these classes with large improvement have similar AP using audio and visual models.

For several sound classes, one of the modalities is much more superior.
One may argue for using only the appropriate single-modal models.
However, we see that in several of these cases the other modality, despite being inferior on its own, can provide complimentary information, which can lead to improved performance through audiovisual learning.
Consider for example, classes such as \emph{Angry Music} and \emph{Plop}, as shown in Figure~\ref{fig:classmap}.
While the audio modality is much more useful for these classes compared to the visual modality, audiovisual learning still leads to considerable improvement as the information from the visual modality compliments the audio modality.
The same inference applies for classes such as \emph{Bus} and \emph{Motorboat}, where the visual modality is much more superior compared to the audio modality. However, there are exceptions to the above observations.
On the audio side, \emph{Whispering} and \emph{Police Siren}, and on the video side, \emph{Shuffling Cards} and \emph{Writing}, are examples of such exceptions.

One final observation we would like to point out from Figure~\ref{fig:classmap} is that, in general for classes with low performance for the audio and visual models (less than 10 AP), audiovisual fusion also does not lead to any improvement.
We believe that a multi-modal framework in which the audio and visual modalities are jointly used to train a single model might be more suitable in these situations.

\subsection{Analysis of Attention}
We analyze the attention weights obtained for each sample for each class.
For each class, we compute the mean attention weight ($\bar{\boldsymbol{\alpha}}_a^c$) by averaging over all evaluation examples which contains the $c^{th}$ class. This gives us an estimate of the importance of the audio modality for the $c^{th}$ class (and consequently for the visual modality as well). 
Figure~\ref{fig:weights} is a histogram plot of the mean attention weights.
The right side of the plot with higher $\boldsymbol{\alpha}_a$ represents dominance of the audio modality, whereas on the left side the visual modality is the dominant modality.
The distribution is clearly skewed towards the audio modality, which is expected since the audio model is better than the visual model for more than 75\% of the sound classes.
There are more than 110 sounds for which on average, the audio modality gets all the weight ($\bar{\boldsymbol{\alpha}}_a^c \approx 1$).
Interestingly, there are around 38 sound classes where the visual modality has complete dominance ($\bar{\boldsymbol{\alpha}}_a^c \approx 0$). Furthermore, for more than 70 sound classes the visual modality on average gets more than 90\% weight ($\bar{\boldsymbol{\alpha}}_a^c \leq 0.1$). This signifies that for several sounds the visual modality can play a vital role in their recognition. 

\section{Conclusions}%
\label{sec:conclusion}

We propose audiovisual models for learning sounds.
Instead of jointly using audio and visual modalities in a single model, we propose to combine the individual models that were trained separately on each modality. To this end, we use a state-of-the-art audio based sound event recognition model and propose a novel vision based model for recognizing sound events. On the Audioset dataset of sound events, it outperforms a prior vision model for sounds by +6.93 mAP (relative: 36.9\%).  
We then propose an attention mechanism for audiovisual fusion that dynamically fuses the outputs from the audio and visual models in a sample-specific and class-specific manner.
Our proposed audiovisual learning model improves state-of-the-art performance on Audioset by approximately +4.35 mAP (relative: 10.4\%).
We also provide thorough analysis of the role of the audio and visual modalities for various sound classes.
The outcomes of this analysis can play a crucial role in the development of future models for audiovisual learning of sounds. 

\bibliographystyle{named}
\bibliography{references}

\end{document}